# SECTOR ANALYSIS
# CHEMICAL INDUSTRY

| | |
|---|---|
| Anandlogesh R R | CH18B037 |
| Breasha Gupta | BE17B015 |
| Divika Agarwal | CH17B043 |
| Rasika Joshi | HS17H011 |

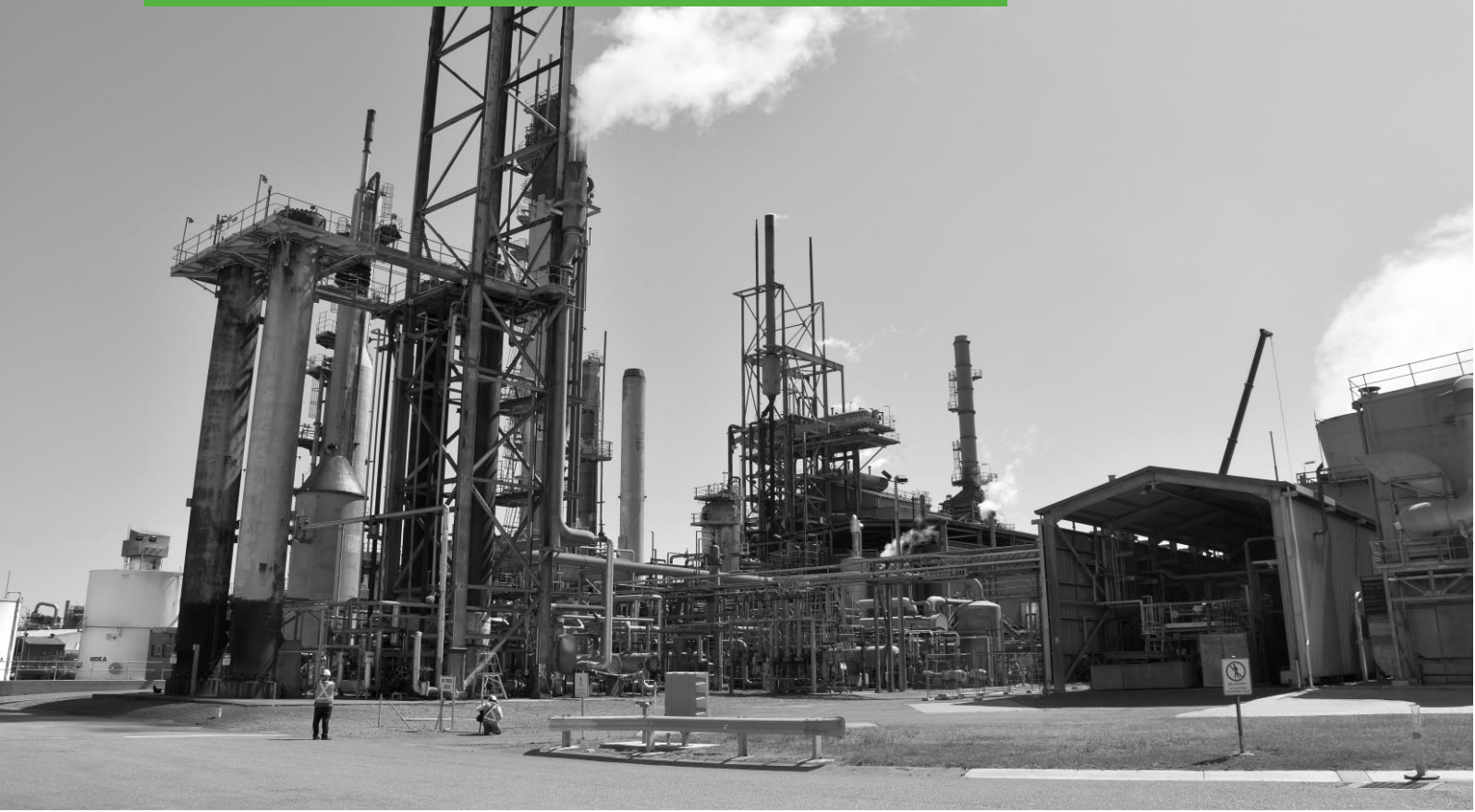

# CONTENTS

# INTRODUCTION

India's chemical story is one of outperformance and promise. A consistent value creator, the chemical industry remains an attractive hub of opportunities, even in an environment of global uncertainty. Worldwide trends affecting the global chemical industry could lead to near-term opportunities for chemical companies in India.

India is an attractive hub for chemical companies. The chemical industry is a global outperformer regarding total returns to shareholders (TRS), and this has resulted in high expectations for sustained, continual growth. The macro perspective on India indicates that while the short-term outlook is challenging, the country's long-term-growth story remains positive.

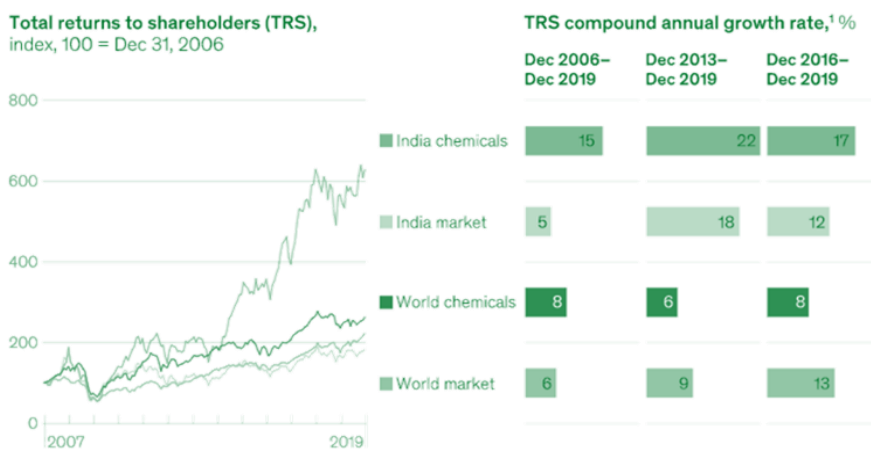

Source[1]

A long-term perspective indicates that India has averaged an annual GDP growth of 7 percent for the last 30 years. The country is also working on becoming a $5 trillion economy. This long-term optimistic scenario bodes well for chemical companies,

---

[1] "Our Insights: India's Chemical Industry" McKinsey
https://www.mckinsey.com/industries/chemicals/our-insights/indias-chemical-industry-unleashing-the-next-wave-of-growth# (Accessed November 19, 2020) especially in light of a long investment cycle. Chemical companies can also benefit from rising domestic demand in chemical end-use sectors, India's attractiveness as a manufacturing destination, and its improved ease of doing business.

# INDUSTRY BREAKDOWN

## SEGMENTATION

The Indian chemical industry can be broadly segmented into the following types based on categories and end-use specification of the chemical produced. The sector is highly diversified with a low Herfindahl index and has deep linkages with the agriculture and manufacturing industries.[2]

1. *Bulk/basic chemicals*

This comprises basic organic chemicals like methanol and acetic acid, and inorganic chemicals like soda ash, chlorine etc. These chemicals are produced through large-scale processes and are typically used as intermediaries in the downstream value chain. There is less opportunity for product-differentiation in this sector, hence the biggest players rely on economies of scale for sustenance. Alkali chemicals constitute almost 70% of total domestic production in this segment.

Tata Chemicals and Gujarat Alkalis and Chemicals Ltd. are the key players in the bulk chemical segment.

2. *Specialty chemicals*

These are low-yield, high-value chemicals that require distinctive technical expertise for production, for example, paints, adhesives, personal care chemicals, food-grade chemicals etc. For this reason, specialty chemicals generate higher EBITDA margins than basic

---

[2] Hait, Angshuman, John, Joice, Das, Abhiman and Mitra, Anujit, "Corporate Pricing Power, Inflation and IIP Growth: An Empirical Investigation", *RBI Working Paper Series No. 09*, 2013, https://www.rbi.org.in/scripts/PublicationsView.aspx?id=15417 (Accessed November 21, 2020)

intermediary products. This segment, which constitutes around a quarter of the total chemical market in India, is majorly geared towards the end-user market and operates at the bottom end of the value chain. Dyes and pigments, flavors and fragrances, surfactants, and personal care products account for more than 80% of the specialty chemical share, with the balance being water, printing, cleaning, and rubber chemicals.

The segment is seeing increased demand in B2B and B2C spaces and is expected to grow at 10% CAGR till 2025 on the back of greater domestic availability of building block intermediaries. Prominent players are Keva, Asian Paints, Pidilite, and Berger.[3]

### 3. Agro-chemicals

This includes 'knowledge chemicals' like fertilizers (nitrogenous, phosphorous, and potassium based) insecticides, pesticides etc., and are produced in bulk amounts. Production of fertilizers for the domestic market is region-specific depending on the variations in agricultural yield and cash crops in the region.

India is the 3rd largest producer of agro-chemicals and exports 50% of its total production. There has been an increase in foreign mergers and acquisitions in this space, notably from global players like DuPont, Dow, and Bayer, leading to a substantial capacity expansion. The risk lies in potential underutilization as a result of this expansion. The segment is expected to grow at a CAGR of 7.5% till 2022 to reach a capitalization of USD 3.7 bn. Key domestic players are Coromandel, Tata subsidy Rallis, and UPL.

### 4. Petro-chemicals

This subgroup refers to chemicals under the olefin class (like ethylene and propylene) and aromatic class (like benzene and xylene isomers). Petrochemicals can be used as intermediaries akin to basic chemical, for example, benzene is used to prepare dyes and ethylene/propylene is used for making plastic products.

---

[3] "Indian Chemical Industry Challenges and Opportunities", *Associated Chambers of Commerce and Industry of India, 2*015, https://www.resurgentindia.com/pro_bfloors/services_img/pdfteders/2093224542Challenges-and-Opportunities-Report-F.pdf (Accessed November 21, 2020)

In India, a key emerging trend is the integration of downstream industries with basic petrochemical refineries. The forecasted growth in demand from FY17 to FY23 is approximately 5% CAGR. GAIL, Haldia, and RIL have long been the dominant firms in this segment.

## 5. Bio-pharma chemicals

This is a small but fast-growing segment that primarily comprises active pharmaceutical ingredients (API), intermediary pharmaceutical products, and xanthine derivatives. APIs include ingredients in drugs for hypertension, cancer, asthma, skin care etc., and are produced from intermediates like perindopril, ter-butylamine, hydrochloride etc. The xanthine derivate space is growing in India due to an influx of energy drinks and specialty beverage products in the market.

The biggest players in the bio-pharma space are also leaders in the larger biotech sector, namely, Piramal Group, GlaxoSmithKline, Krebs Biochemicals and Industries Ltd. (KBIL).

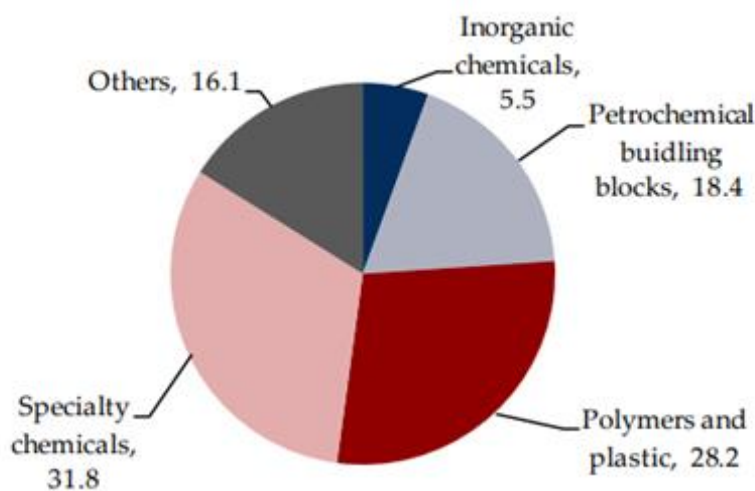

India sector-wise exports in FY19 (%)[4]

---

[4] "Chemical Industry Sector Update", *HDFC Investment Advisory Group,* May 7, 2018
https://v1.hdfcbank.com/assets/pdf/privatebanking/SectorUpdate-Chemical-07052018.pdf (Accessed November 21, 2020)

# VALUE CHAIN

The general production process of chemicals can be understood through the following flowchart. Value chains are chemical-specific, for example, centered around chlorine, hydrogenation, ATBS etc.

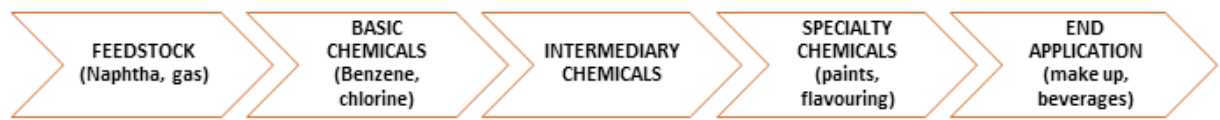

There are two trends identified among firms operating at different levels along this value chain. First, some firms choose to stay contained within a single level (such as feedstock or bulk chemicals) and take up horizontal integration within that category. Second, some players opt for upstream or downstream integration with a view towards product-differentiation across multiple baskets.[5]

Moving upstream in the value chain allows a firm to take control of its raw materials and intermediaries. It does not have to rely on bulk suppliers, thus saving transaction costs and eliminating market contingencies to a large extent. However, while operating in the basic chemicals space, margins are low and the firm needs to produce at scale to sustain its upstream integration.

Moving downstream entails a foray into the end user market by producing high-value added specialty chemicals. Due to stringent IPR and technical expertise required in this section of the value chain, downstream integration is a long-term and capital-intensive process.

# FOCUS AREAS

Focus areas that will drive growth in the short to medium term are:

❖ *Growing domestic consumption* – Import substitution models for the domestic market

❖ *Increasing mergers and investments from multinational companies* - The government has taken various steps like allowing 100% FDI in the Chemical sector, removed licensing requirements except in the case of hazardous chemicals, etc. in order to increase FDI inflow into the sector.

---

[5] "Spurting the growth of Indian Chemical Industry: Handbook on Indian Chemicals and Petrochemicals Sector", *FICCI,* October 2014, http://ficci.in/spdocument/20441/Knowledge-Paper-chem.pdf (Accessed November 21, 2020)

The sector has seen massive FDI inflow over the last four years, mainly due to improvement in utilization levels in the industry, which has called for fresh investment demand.

- ❖ *Exploring alternative feedstock options* – Cheaper methods of naphtha cracking are making the feedstock segment more competitive across the country. The depreciation in the price of oil also supports the increase in capacity and margins of naphtha exports. India's agricultural base can be leveraged for biomass feedstock though bio methanation processes.

- ❖ *International macro trade developments* – The US-China trade war could present an opportunity for Indian players to step by exports and shift the global supply chain away from China.

- ❖ *Environmental laws in China and EU* – Following stricter environmental regulations in the Chinese and European chemical sector, Indian bulk and intermediary manufactures have an opportunity to benefit from the filling the space. [6]

---

[6] Raghavan, Ravi and Joshi, J.B., "India's Expanding Chemical Industry", *American Institute of Chemical Engineers,* Dec 2018, https://www.aiche.org/sites/default/files/cep/20181249.pdf (Accessed November 21, 2020)

# KEY PLAYERS

## AARTI INDUSTRIES LTD.

**QUALITATIVE ANALYSIS**

- ❖ Revenue - Product Mix:

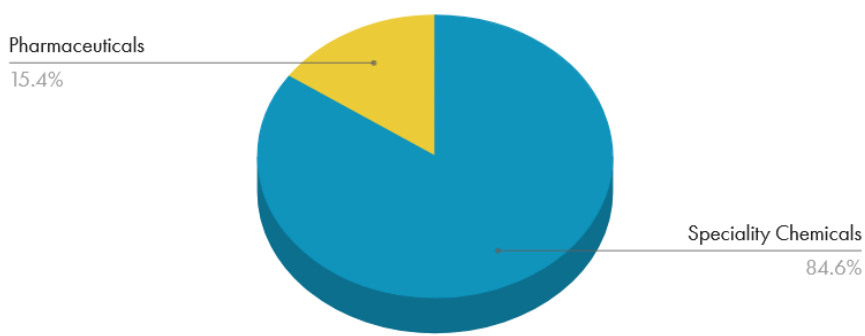

[Source](#)

### *Speciality Chemicals:*

❖ Global Leader

The company is the global market leader in the manufacturing of benzene-based derivatives

| Product | Global Rank |
| --- | --- |
| Chlorination | Top 3 |
| Nitration | Top 3 |
| Ammonolysis | Top 2 |
| Hydrogenation | Top 2 |
| Halex Chemistry | Only Player in India |

### *Pharmaceuticals:*

Develop APIs, intermediates and xanthine derivatives for the pharmaceutical and food/beverages industry.

- US Food and Drug Administration (FDA) and EU GMP (European Union Goods Manufacturing Practice) accreditation

- Intellectual Property Rights (IPR) support for global markets

## *Overall:*

❖ Well-diversified
- In terms of end-user industries: agrochemical, pharmaceuticals, polymers and additives, and dyes and pigments. Revenue contribution from each of these end-user industries is ~15-25%
- In terms of geographic distribution:

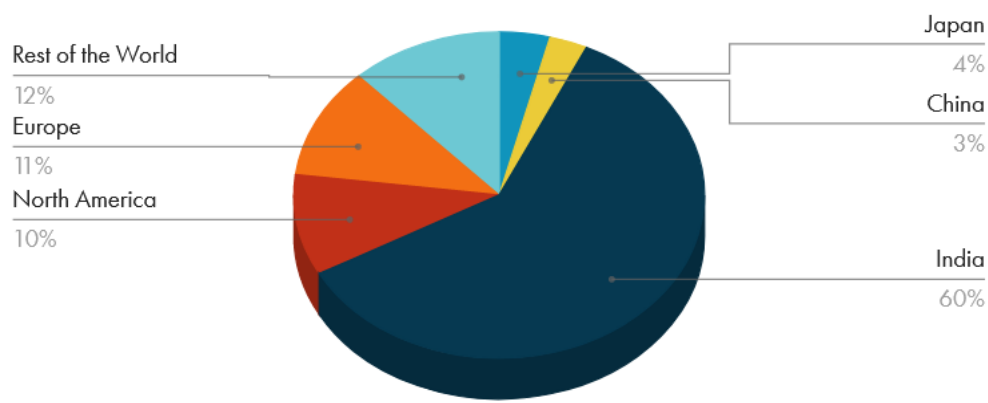

[Source](#)

- More than 400 international and 700 domestic customers
- Its customer base is well-diversified with top 10 and 20 customers contributing 27% and 38% respectively to revenues, while the biggest customer contributes ~4% only.

❖ Strong supply chains
- Highly internally integrated across the product chains of benzene and toluene
- Provides a formidable threat to new entrants

❖ Investment in R&D and Capacity Growth
- The company has invested in expanding the capacity of Nitrochlorobenzene and Phenylenediamine to cater to growing end-user industry demands.
- The company is also venturing into Toluene chemistry space. The toluene segment in India is untapped and catered mainly through imports; hence AIL will benefit immensely by entering this segment.

- The company has set up its 4th R&D plant in Q4FY20; providing 2 plants dedicated to Speciality Chemicals and Pharmaceuticals each. With a healthy growth rate 42% CAGR in R&D expenditure (over FY16-20) in strategic spaces, the company is considered forward looking in its business.[7]

## FINANCIAL ANALYSIS (Consolidated)

### P&L:

| (In INR mn) | FY18 | FY19 | FY20 |
|---|---|---|---|
| Revenue | 38,061 | 47,055 | 41,863 |
| EBITDA | 6,991 | 9,651 | 9,773 |
| PBT | 4,290 | 6,220 | 6,762 |

### Financial Ratios:[8]

|  | FY18 | FY19 | FY20 |
|---|---|---|---|
| Quick Ratio | 0.59 | 0.93 | 0.61 |
| Debt Equity Ratio | 1.32 | 0.91 | 0.70 |
| Asset Turnover Ratio | 0.87 | 0.71 | 0.66 |
| ROCE (%) | 19.99 | 20.49 | 18.14 |
| Operating Profit Margin (%) | 18.57 | 23.21 | 23.56 |

The company has been able to earn a healthy return on its capital in the midst of the Capex that it has incurred. ROCE will fall in the coming few years (FY 21-22) due to the fall in EBIT and investment of Capex that that takes time to reap results.

---

[7] Aarti Industries Ltd., "Annual Report 2019-2020",
https://www.aarti-industries.com/media/investors/annual/1598586177_Annual_Report_2019-2020.pdf
(Accessed November 15, 2020)

[8] Tijori, "Aarti Inds: Tijori Finance."
https://www.tijorifinance.com/company/aarti-industries-limited. (Accessed November 15, 2020)

# AKLYL AMINES CHEMICALS LTD.

## QUALITATIVE ANALYSIS

- ❖ Revenue – Product Mix:

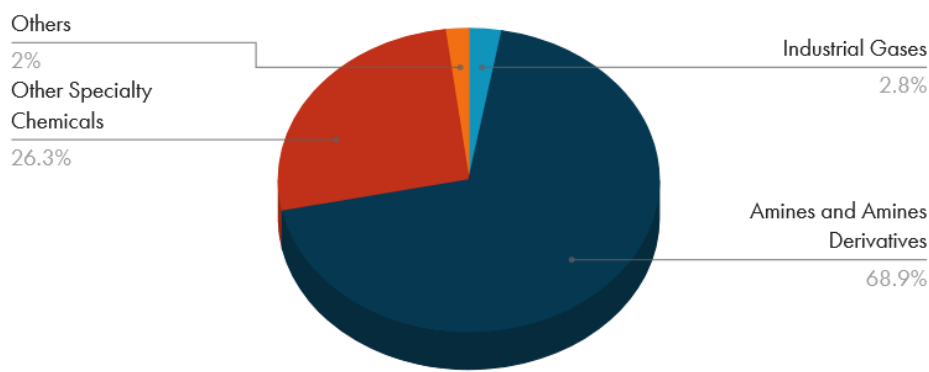

[Source](#)

- ❖ Revenue – End-user Industry Mix:

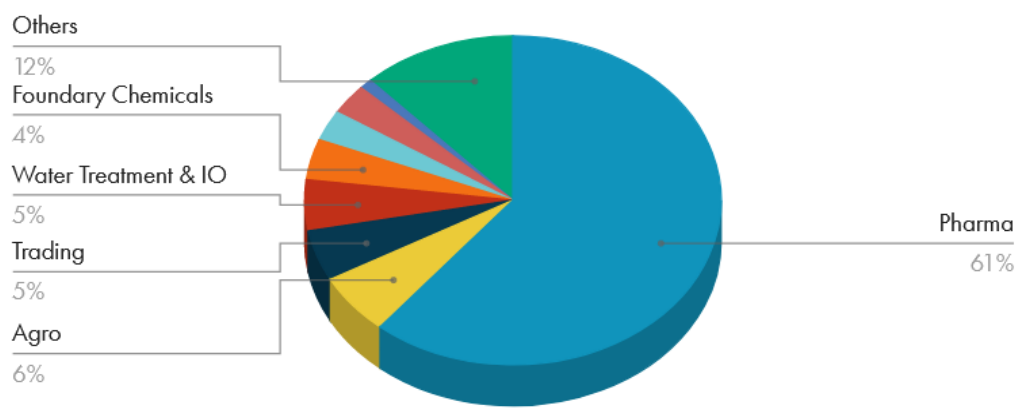

[Source](#)

- ❖ Positive outlook on Growth of End-user Industries
    - Since the company is in the business of manufacturing and supplying pharmaceutical intermediates, its volume growth is directly proportional to that of the global pharma and agrochemical industries.
    - Global spending on pharmaceuticals is set to grow at a 3-6% CAGR over CY19-24 to USD 1.5-1.6 trillion.[9]

---

[9] Aurobindo Pharma, "Annual Report 2019-20",

- ❖ Future Capacity Augmentation Plans
    - The company plans to spend a total of INR 3bn over FY20-22E to expand its
        - Methyl Amines (and its derivatives) capacity by 15ktpa to 45 (timeline 3QFY21)
        - Acetonitrile capacity by 15ktpa to 27 (timeline 2QFY22)
    - At peak utilisation, both plants put together are expected to contribute INR 4-5bn to the topline at current prices.[10]

## FINANCIAL ANALYSIS (Consolidated)

*P&L:*

| (In INR mn) | FY18 | FY19 | FY20 |
|---|---|---|---|
| Revenue | 6,162 | 6,162 | 6,162 |
| EBITDA | 8,464 | 8,464 | 8,464 |
| PBT | 958 | 1,302 | 2,507 |

*Financial Ratios:[11]*

|  | FY18 | FY19 | FY20 |
|---|---|---|---|
| Quick Ratio | 0.85 | 0.83 | 1.37 |
| Debt Equity Ratio | 0.61 | 0.45 | 0.16 |
| Asset Turnover Ratio | 0.86 | 0.98 | 1.08 |
| ROCE (%) | 19.55 | 23.70 | 31.15 |
| Operating Profit Margin (%) | 19.65 | 19.88 | 26.59 |

---

https://www.aurobindo.com/wp-content/uploads/2020/08/Aurobindo-Pharma-Limited-Annual-Report-2019-20.pdf (Accessed November 16, 2020)

[10] Alkyl Amines Chemicals Ltd., "Annual Report for the year 2019-20", http://alkylamines.com/investors/annualreport_1920.pdf (Accessed November 16, 2020)

[11] Tijori, "Alkyl Amines Chem | Tijori Finance."
https://www.tijorifinance.com/company/alkyl-amines-chemicals-limited (Accessed November 16, 2020)

The company maintains a good record of Efficiency and Profitability Ratios. With the implementation of the expansion plans, the company is set to reap benefits for the investors for years to come.

# BALAJI AMINES LTD.

## QUALITATIVE ANALYSIS

- ❖ Revenue - Product Mix:

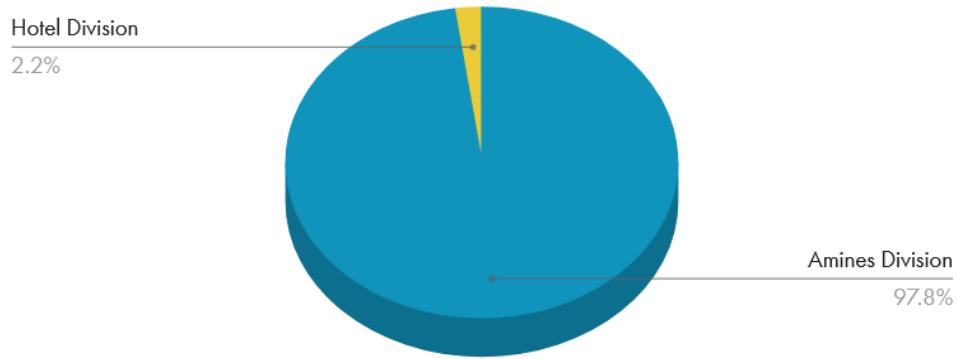

[Source](#)

- ❖ Revenue – End-user Industry Mix:

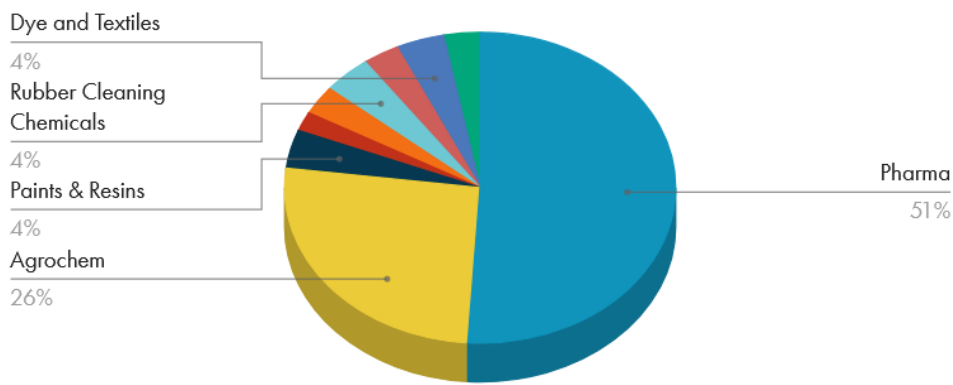

[Source](#)

- ❖ Positive outlook on the End-user Industry
  - Balaji Amines' aliphatic amines volumes grew by 6% YoY in FY20 to 65kT, spurred by growth in demand in the pharma and agrochemical sectors. These together make 75-80% of the customer mix for BLA.

- ❖ Future Capacity Augmentation Plans
  - BLA is presently undertaking a greenfield expansion on a 90-acre land in Solapur

- After a debottlenecking drill from Nov-2020, production levels of Acetonitrile will double from the current 9 ton/day to 18

❖ Investment in Loss Making Businesses
- Balaji Amines has burnt capital in unrelated and low margin businesses such as its hotel business (started in FY14; investment INR 1.1bn) and Compact Fluorescent Lamp (acquired via amalgamation in FY18).
- With the COVID-19 wreaking havoc on the Hotel Industry, the business unit is unlikely to make profits till FY22.[12]

## FINANCIAL ANALYSIS (Consolidated)

### P&L:

| (In INR mn) | FY18 | FY19 | FY20 |
|---|---|---|---|
| Revenue | 8,612 | 9,431 | 9,358 |
| EBITDA | 1,895 | 1,934 | 1,807 |
| PBT | 1,658 | 1,651 | 1,311 |

### Financial Ratios:[13]

|  | FY18 | FY19 | FY20 |
|---|---|---|---|
| Quick Ratio | 0.94 | 0.99 | 1.35 |
| Debt Equity Ratio | 0.33 | 0.40 | 0.39 |
| Asset Turnover Ratio | 1.02 | 0.90 | 0.84 |
| ROCE (%) | 29.48 | 24.15 | 18.16 |
| Operating Profit Margin (%) | 22.48 | 20.96 | 19.85 |

The investment in loss making business units has made the EBITDA margin fluctuate over the years. However, there is a light of hope with the unabated growth in demand from its pharmaceutical and agrochemical customers that formed ~77% of BLA's 1QFY21 revenue mix.

---

[12] Balaji Amines Ltd., "Balaji Amines Ltd. 32nd Annual Report 2019-2020",
http://www.balajiamines.com/pdf/1594640113AnnualReport2019-20.pdf (Accessed November 16, 2020)

[13] Tijori, "Balaji Amines | Tijori Finance."
https://www.tijorifinance.com/company/balaji-amines-limited (Accessed November 16, 2020)

# SRF

## QUALITATIVE ANALYSIS

❖ Revenue – Product Mix:

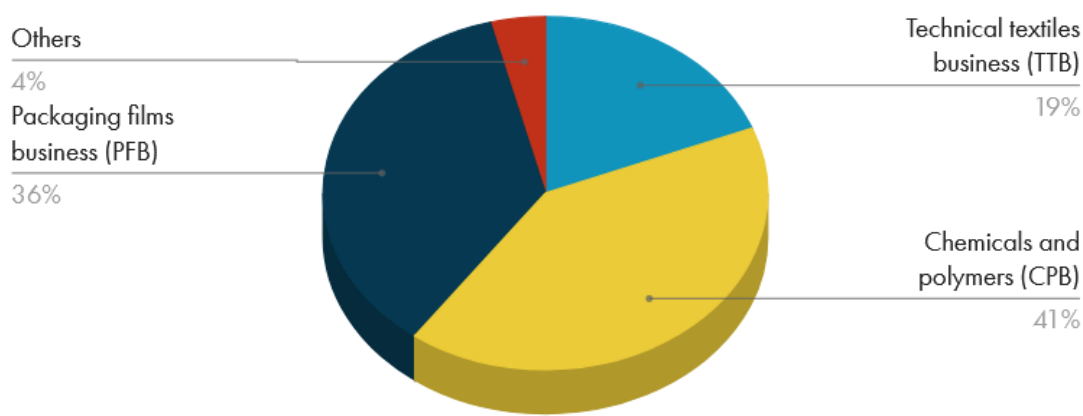

## *Chemicals*

The segment comprises two sub-segments mainly, fluorochemicals and speciality chemicals. The fluorochemicals sub-segment derives its revenue from (1) refrigerant gases, and (2) chlorinated solvents and industrial chemicals. Revenue of fluorochemicals grew at a CAGR of 11.6% from FY11 to FY20.[14]

Speciality chemicals cater to agrochemical (80%) and pharmaceutical industries (20%). Revenue of speciality chemicals grew at a CAGR of 37% from FY11 to FY20 which resulted in a jump in its contribution to the segmental revenue from 13% in FY11 to 55% in FY20.[15]

❖ Positive outlook on Chemical Industry
- The role of fluorine in drug design and development is expanding rapidly owing to its membrane permeability, metabolic pathways, and pharmacokinetic properties. SRF has developed an expertise in fluorination chemistry and is in the sweet spot to grab this opportunity.

---

[14], [14] Tijori, "SRF | Tijori Finance."
https://www.tijorifinance.com/company/srf-limited (Accessed November 16, 2020)

- SRF has already developed and supplied fluoro speciality intermediates for various customers in pharmaceutical and agrochemical industries. In FY20, the company has launched six new agro intermediates and three pharmaceutical intermediates.

❖ Capacity Augmentation Plans
- The company continues to focus on R&D and has plans to launch 3-4 new products every year from which the company will be able to deploy dedicated facilities for some of them. Steady capacity expansion, increasing R&D spend, and an expanding customer base will result in sustaining its healthy growth momentum.

❖ Alternatives for Refrigerants
- R22 is being phased out from developing countries with an expected 67.5% reduction in supplies by 2025 and a total ban on the refrigerant gas is expected by 2030 as per international regulations.
- The company is deploying two ways to tackle this challenge:
    - Developing drop-in substitutes for R-22, and
    - Use of R-22 as a raw material.

## *Packaging Films*

Packaging films business (PFB) grew from INR 8.7bn in FY11 to INR 26bn in FY20 at a CAGR of 13%. Revenue contribution from this segment has increased from 25% in FY11 to 36% in FY20.

❖ Investment in Capacity Expansion
SRF has commissioned a 40ktpa BOPET plant each in Thailand and Hungary in 1QFY21 and 2QFY21 respectively.

The overall segmental growth will be driven by a favourable situation in global BOPET market and recently added capacities by the company.

## *Textile Business*

SRF is the largest manufacturer of nylon tyre cord fabrics (NTCF) in India and the second-largest manufacturer in the world. It is the second-largest manufacturer of conveyor belting fabrics in the world.

This business segment has significant exposure to the tyre industry. The tyre cord fabric, which is extensively used in automobile tyres, contributes majorly to the segmental revenue.

However, the slowdown in the automobile industry led to a de-growth of a 12.3% CAGR over FY17-20. Thus, the segmental contribution to the overall revenue has declined from 53% in FY11 to 19% in FY20. from this segment has increased from 25% in FY11 to 36% in FY20. [16]

## FINANCIAL ANALYSIS (Consolidated)

*P&L:*

| (In INR mn) | FY18 | FY19 | FY20 |
|---|---|---|---|
| Revenue | 55,108 | 69,499 | 70,621 |
| EBITDA | 9,062 | 12,970 | 14,549 |
| PBT | 5,817 | 7,684 | 9,147 |

*Financial Ratios:[17]*

|  | FY18 | FY19 | FY20 |
|---|---|---|---|
| Quick Ratio | 0.58 | 0.61 | 0.52 |
| Debt Equity Ratio | 0.88 | 0.90 | 0.82 |
| Asset Turnover Ratio | 0.67 | 0.78 | 0.68 |
| ROCE (%) | 12.10 | 15.33 | 16.83 |
| Operating Profit Margin (%) | 18.27 | 17.98 | 22.37 |

Capacity expansion and the plan to launch 3-4 new products every year will sustain growth momentum. This could, however, impact the ROCE in the FY21 season.

---

[16] SRF, "Annual Report FY2019-20",
https://www.srf.com/pdf/annual-reports/FY%202019-20/SRF%20AR%202019-20_F1.pdf (Accessed November 16, 2020)

[17] Tijori, "SRF | Tijori Finance."
https://www.tijorifinance.com/company/srf-limited (Accessed November 16, 2020)

# NAVIN FLUORINE

## QUALITATIVE ANALYSIS

❖ Revenue - Product Mix:

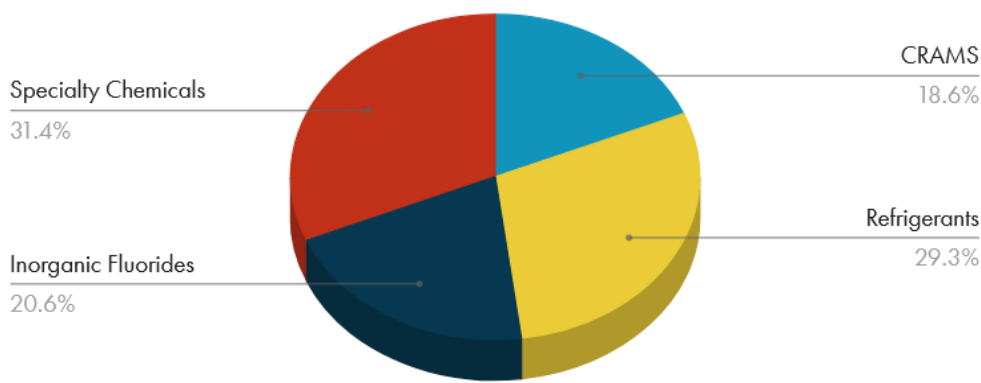

[Source](#)

❖ Pharma and Agrochemical industries lead the growth

- Over FY17-20, EBITDA margins have remained over the 20% mark, led by robust demand from pharma and agrochemical customers for speciality chemicals as well as higher traction in the Contract Research and Manufacturing Services (CRAMS) Business Unit.
- Unabated demand from pharma/agrochemical customers can lead to a tilt in product mix towards high-margin high-value business.
- The growth in speciality chemicals' revenue was backed by product portfolio expansion, deeper penetration into existing users and robust project pipeline in life/crop science.

❖ Future Capacity Augmentation Plans

- Ramp-up in the Certified Good Manufacturing-3 (cGMP-3) plant boosted revenues for the CRAMS BU.
- NFIL will be investing USD 51.5mn (~INR 3.7bn) in setting up the manufacturing facility and ~USD 10mn (INR 0.7bn) for a captive power plant. The augmented capacity will be operational from 4QFY22.[18]

---

[18] Navin Fluorine International Ltd., "Annual Results in Financial Year 2019-20",
[https://www.nfil.in/investor/financial_results/fy_2019_2020/annual_report_for_the_financial_year_2019_20.pdf](https://www.nfil.in/investor/financial_results/fy_2019_2020/annual_report_for_the_financial_year_2019_20.pdf)
(Accessed November 20, 2020)

# FINANCIAL ANALYSIS (Consolidated)

*P&L:*

| (In INR mn) | FY18 | FY19 | FY20 |
|---|---|---|---|
| Revenue | 9,127 | 9,959 | 10,616 |
| EBITDA | 2,150 | 2,184 | 2,635 |
| PBT | 2,665 | 2,244 | 2,578 |

*Financial Ratios:[19]*

|  | FY18 | FY19 | FY20 |
|---|---|---|---|
| Quick Ratio | 1.62 | 1.64 | 3.53 |
| Debt Equity Ratio | 0.01 | 0.00 | 0.00 |
| Asset Turnover Ratio | 0.56 | 0.57 | 0.53 |
| ROCE (%) | 21.65 | 17.02 | 14.28 |
| Operating Profit Margin (%) | 33.69 | 25.38 | 27.96 |

The company is virtually debt free – with the industry low Debt Equity Ratio. However, reducing ROCE and low Asset Turnover Ratio is a cause of concern for the shareholders, which, if rectified provides an amazing investment opportunity.

---

[19] Tijori, "Navin Fluorine Intl | Tijori Finance."
https://www.tijorifinance.com/company/navin-fluorine-international-limited (Accessed November 20, 2020)

# VINATI ORGANICS

## QUALITATIVE ANALYSIS

- ❖ Revenue - Product Mix:

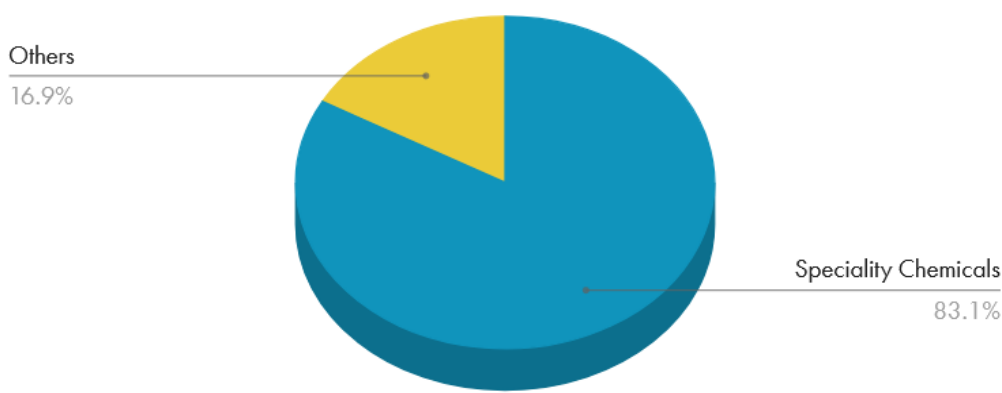

[Source](#)

VO is the world's largest manufacturer of Iso Butyl Benzene (IBB) and 2-Acrylamido 2-Methylapropane Sulphonic Acid (ATBS) and holds 65% of the global market share for both the products (FY19). 76% of the revenue mix in FY19 was derived from these two products alone.

- ❖ Falling ATBS Demand
  - ATBS finds application in water treatment, construction, personal care, among others; the oil and gas industry forms 25-30% of the customer mix for ATBS.
  - With the fall in crude prices in CY20, the demand for ATBS by the oil customers took a hit as well, which was amply evident from the fall in contribution of the product in the 1QFY21 revenue mix to 40% from 57% in FY20.

- ❖ Rising IBB Demand
  - IBB is a basic raw material of the Ibuprofen bulk drug.
  - On the other hand, the demand of IBB has risen in tandem with the 25% rise in the global Ibuprofen demand over FY16-20 (evident from capacity expansion by 20/33% for Ibuprofen/IBB by an Indian competitor in FY19); the contribution of IBB in the revenue mix shot up to 32% in 1QFY21 from 16% in FY20.[20]

---

[20] *HDFC Securities, 2*020, https://trendlyne.com/research-reports/stock/1485/VINATIORGA/vinati-organics-ltd/ (Accessed November 20, 2020)

Since ATBS is a higher margin product, the value of EBITDA might take a toll for the foreseeable future.

- ❖ Capacity Augmentation plans:
  - The Butyl Phenol plant (capacity 35ktpa) had started operating from Sep-19, but its ramp-up has been slow in FY20. The business would primarily be driven by domestic sales (60-70% of installed capacity) as it substitutes for the more expensive imports, while the export market would take 2-3 years to pick up.
  - The brownfield expansion of the ATBS capacity by 14ktpa to 40, costing INR 1.1bn, is likely to come on stream by 4QFY21 as the demand for the product should improve.[21]

## FINANCIAL ANALYSIS (Consolidated)

*P&L:*

| *(In INR mn)* | FY18  | FY19   | FY20   |
|---------------|-------|--------|--------|
| Revenue       | 7,297 | 11,081 | 10,289 |
| EBITDA        | 1,973 | 4,036  | 4,139  |
| PBT           | 2,034 | 4,252  | 4,247  |

*Financial Ratios:*

|                            | FY18  | FY19  | FY20  |
|----------------------------|-------|-------|-------|
| Quick Ratio                | 3.83  | 4.94  | 6.17  |
| Debt Equity Ratio          | 0.33  | 0.40  | 0.39  |
| Asset Turnover Ratio       | 0.74  | 0.92  | 0.71  |
| ROCE (%)                   | 0.02  | 0.00  | 0.00  |
| Operating Profit Margin (%)| 31.37 | 40.32 | 44.70 |

With the above listed problems against Vinati Organics, the company might struggle to find good sources of income. But the shift in focus towards IBB gives a light of hope.

---

[21] Vinati Organics Ltd., "Annual Report for the Financial Year April 2019 To March 2020", https://vinatiorganics.com/uploads/7280VINATI%20ORGANICS%20LTD%20ANNUAL%20REPORT%2031.03.2020.pdf (Accessed November 20, 2020)

# TRENDS GOING FORWARD

Global trends that are shaping the near-term future of the Indian chemical industry:[22]

1. Major global oil and gas companies are now shifting focus towards the downstream petroleum industry. Following suit on this trend India may see higher investments in the petrochemical sector. This could drive the industry towards self-sufficiency.

2. Digitalization and technology have in general gotten a huge boost in the last couple of years and is even more heightened due to COVID-19. Embracing the possibilities of digital transformation in this industry could help Indian companies increase efficiency and expand their profit margins.

3. Every industry globally is looking at sustainability much more seriously. With the Indian chemical industry various stakeholders are also willing to pay a premium for this. Environment sustainability will become a driving factor for enhancing long term shareholder value and being in compliance with the government regulations

4. Industry wide a trend of focusing on the core business while consolidating all players is being seen. The scale at which companies choose to operate will define competitive edge especially in a country like India and mergers and acquisitions to consolidate businesses is a future trend that has already begun.

5. The ongoing trade disputes amongst major players like US and China and India and China are making the entire industry highly unpredictable and directly affects investor sentiments and that could be a potential threat to the industry[23]

6. The chemical industry is closely related to the major end markets like automotive and construction. Abroad both these industries have taken a huge hit and the trend is following suit

---

[22] "Our Insights: India's Chemical Industry" McKinsey https://www.mckinsey.com/industries/chemicals/our-insights/indias-chemical-industry-unleashing-the-next-wave-of-growth# (Accessed November 19, 2020)

[23] "2020 Chemical Industry Outlook" Deloitte – https://www2.deloitte.com/us/en/pages/energy-and-resources/articles/chemical-industry-outlook.html (Accessed on November 18, 2020)

in India with increased spending towards electric vehicles and used cars. These trends will actually see a shift within the chemical industry as they invest more on innovation across new products, technology and business models

# CONCLUSION

The Indian chemical industry is speeding ahead, especially the new trend of consumption and production shifting to South Asian countries across all sectors has drastically increased the demand for chemicals and petrochemicals. However, it still comes with its set of issues that need to be resolved. The government needs concrete permit approval processes that are in tandem with global standards of environment exposure and pollution norms. With innovation being the key trend in this industry, piracy and counterfeiting issues will have to be dealt with urgently if the country wants to set up an efficient, transparent and flexible environment for these innovators.[24]

Another thing to be focused on by the government is boosting foreign direct investment opportunities for these industries. In spite of 100% FDI being allowed under the automatic route in this sector other than hazardous chemicals, we only see foreign investments comprising 4% of total investment in this Indian sector.

Some of the key success drivers for this industry going forward could include:
- ❖ Proximity to strong growth markets
- ❖ Increase of manufacturing within the country
- ❖ Greater ease of doing business
- ❖ Continued development of PCPIRs

Based on all the above trends and success drivers it is estimated that the key growth segments of this industry are: Petroleum and petrochemicals, specialty chemicals, chlor alkali, pesticides, pharmaceuticals and bulk drugs.

---

[24] "Rise of Indian Chemical sector" Process Worldwide
https://www.process-worldwide.com/rise-of-the-indian-chemical-sector-a-981402/
(Accessed November 19, 2020)

Today India's chemical sector is growing one of the fastest globally but in order to compete globally our chemical industry needs a huge boost from the government with the help of policies and grants because today majority of our sector is still comprised of small to midsize businesses who can't afford to spend on world class technologies, whereas the technology being used by bigger medium to large businesses are coming from overseas. Indian chemical companies spend only 1% on R&D whereas the global benchmark is 5% and this can be leveraged on by higher investments in innovation to compete on a global scale.